# Long-term deep supercooling of large-volume water via surface sealing with immiscible liquids

Haishui Huang[1], Martin L. Yarmush[1,2]*, O. Berk Usta[1]*

[1] Center for Engineering in Medicine, Massachusetts General Hospital, Harvard Medical School and Shriners Hospitals for Children, Boston, Massachusetts 02114, United States
[2] Department of Biomedical Engineering, Rutgers University, Piscataway, New Jersey 08854, United States

* Correspondence should be addressed to:

O. Berk Usta, PhD
Center for Engineering in Medicine (CEM)
Massachusetts General Hospital
Harvard Medical School
Shriners Hospitals for Children
51 Blossom Street
Boston, MA 02114
E-mail: berkusta@gmail.com

Martin L. Yarmush, MD, PhD
Center for Engineering in Medicine (CEM)
Massachusetts General Hospital
Harvard Medical School
Shriners Hospitals for Children
51 Blossom Street
Boston, MA 02114
E-mail: ireis@sbi.org

*Abstract:*

*Supercooling of aqueous solutions below their melting point without any crystallization is a fundamentally and practically important physical phenomenon with numerous applications in biopreservation and beyond. Under normal conditions, heterogeneous nucleation mechanisms critically prohibit the simultaneous long-term (> 1 week), large volume (> 1 ml) and low temperatures (< -10 $^{o}C$) supercooling of aqueous solutions. Here, in order to overcome this bottleneck and enable novel and practical supercooling applications, we report on the use of surface sealing of water by an oil phase to drastically diminish the primary heterogeneous nucleation at the water/air interface. Using this approach, we have achieved deep supercooling (as low as -20 $^{o}C$) of large-volumes of water (up to 100 ml) for long periods (as long as 100 days) simultaneously. Since oils are mixtures of various hydrocarbons we also report on the use of pure alkanes and primary alcohols of various lengths to achieve the same. All alcohols and some of the longer alkane chains we studied show high capacity to inhibit freezing. The relationship of this capacity with the chain length, however, shows opposite trends for alcohols and alkanes due to their drastically different interfacial structures with the water molecules. We find that the deeply supercooled water (at -20 $^{o}C$) can withstand vibrational and thermal disturbances with all sealing liquids used, and even an extreme disturbance, ultrasonication, when alcohols are used as the sealing phase. The deep supercooling approach we developed here, for large samples and long periods, is expected to enable novel applications of supercooling in a variety of areas including biopreservation and food storage among others.*

Water is a seemingly simple yet a practically complex liquid with extraordinary phase behavior, which enables many of life's intricacies. While water is possibly the most studied liquid, there remain many areas where its behavior is still largely mysterious (*1*). A prime example for this is the freezing and the supercooling of water, one of the most important yet least understood phenomena in our daily lives and scientific research (*2, 3*). Ice formation and the preceding supercooled state of microdroplets in atmospheric clouds is a crucial element for precipitation and reflection of solar radiation (*4, 5*). Furthermore, chilling, freezing, freeze avoidance, and supercooling are important strategies to combat cold weather for ectothermic animals (*6, 7*), treat malignant diseases via cryotherapy (*8*), and preserve food and various biological samples, such as cells, tissues, and organs (*9, 10*).

Recent advances have shown that supercooling can be a promising alternative approach for the preservation of cells, tissues, and especially organs (*11*). Nevertheless, an important hurdle for supercooling preservation, as well as other applications of supercooling, is that simultaneous low temperature (< -10 °C), large volume (> 1ml), and long period (> 1 week) of supercooling for aqueous solutions cannot be readily achieved (*12-14*). While high pressure based approaches have provided supercooled states of water down to -92 °C briefly (*1*), according to the water phase diagram.They are, however, expensive, might further complicate preservation of biological samples, and their long-term fate is unknown. Few experiments have unstably supercooled large volumes, several hundred milliliters, of water to -12 °C (*15*), albeit also for periods on the order of seconds. Similarly, in Dorsey's classical work on freezing of supercooled water, he was able to achieve a temperature of -19 °C for a few milliliters of water very briefly during his constant cooling experiments (*16*). A method that overcomes these hurdles and enable long-term supercooling of large aqueous samples at low temperatures could enable applications in biopreservation as well as many other areas which have previously been practically prohibited.

Under normal atmospheric conditions, ice transitions into water at 0 °C, i.e. the melting point or the equilibrium temperature ($T_e$). Nevertheless, the observed freezing temperature ($T_f$) for pure water is usually below the equilibrium temperature ($T_e$). Water, in the liquid phase, below the equilibrium temperature is said to be "supercooled" where $\Delta T = T_e - T_f$ measures the degree of supercooling. Supercooled water is intrinsically metastable and can spontaneously transform to lower-energy-level ice crystals through the formation of ice nuclei, which can be readily achieved by ice seeding (*17*), ultrasonicating (*18*), or presenting ice-nucleating agents (*19*). On the contrary, it is very difficult to maintain supercooled water unfrozen, especially for a large volume, under a high degree of supercooling ($\Delta T$), or for a long period, as each of these increases the possibility of ice nucleation and water freezing (Supplemental Information (SI)). For instance, $\Delta T$ of a water droplet decreases logarithmically with increasing volume under a constant cooling rate (*20*). Similarly, supercooling frequency ($f_s$, $f_s$ = number of unfrozen droplets / number of total droplets) of an ensemble of droplets decreases exponentially with increasing droplet volume, storage time, and nucleation rate ($J$) (*21, 22*), while $J$ itself increases exponentially with $\Delta T$ (*23*). Consequently, simultaneous long-term (> 1 week), large-volume (> 1 ml), and deep supercooling (DSC, $\Delta T > 10$ °C) of water has not yet been achieved. In what follows, we describe a novel and unexpected method based on sealing of the water surface by an immiscible hydrocarbon based liquid, such as oils, pure alkanes and pure primary alcohols. This method, as we demonstrate through a series of experiments, enables stable supercooling of large volumes of water for long periods at temperatures well below -10 °C.

There are two general ice nucleation mechanisms, homogenous and heterogeneous crystallization. Homogeneous crystallization occurs due to random aggregation of interior water molecules to create a critically large nucleus of ice crystal, which could only be achieved and observed below -20 °C (*24*). Heterogeneous crystallization, on the other hand, stems from ice nucleus formation catalyzed by a substrate and/or with the aid of foreign objects at much higher temperatures (*25*). Consequently, water freezing is generally initiated by heterogeneous nucleation, and especially, the water/air interface is the primary nucleation site as revealed in theoretical (*26, 27*), experimental (*14, 28*), and numerical (*29, 30*) studies. When water molecules aggregate on the water surface (water/air interface) to form an ice nucleus, they need to overcome an energy barrier $\gamma^{ia} - \gamma^{wa}$ ($\gamma$: interfacial tension, superscripts $i, w, o$, and $a$ refer to ice, water, oil, and air, respectively) per unit area as the ice/air interface replaces original water/air interface. In comparison, the energy barrier for homogeneous ice nucleation within bulk water is proportional to the water/ice interfacial tension, $\gamma^{wi}$. This interfacial tension can be expressed via the Young equation as $\gamma^{wi} = \gamma^{ia} - \gamma^{wa} cos\theta_{iwa} \geq \gamma^{ia} - \gamma^{wa}$ ($\theta_{iwa}$: water contact angle on ice/water/air interface, Fig. S1(A)). This inequality indicates that heterogeneous ice nucleation on the surface is thermodynamically more favorable than homogeneous nucleation in bulk as complete wetting ($\theta_{iwa} = 0°$) is not generally observed (*27*), and receding contact angles of 12° have been reported (*31*). Therefore, if the water surface is sealed by an oil phase, the energy barrier of ice nucleation at the water-oil interface would be $\gamma^{io} - \gamma^{wo}$. Similarly, the homogenous nucleation energy barrier can be now expressed in terms of another triple interface, namely the oil/water/ice as $\gamma^{wi} = \gamma^{io} - \gamma^{wo} cos\theta_{iwo}$ where $\theta_{iwo}$, for many oils can be nearly 0° (Fig. S1(B)) (*28, 32*). In the case of $\theta_{iwo} \cong 0$, the energy barrier approaches the limiting case $\gamma^{io} - \gamma^{wo} \cong \gamma^{wi}$. This analysis indicates that the energy barrier of heterogeneous crystallization at the surface is elevated almost to the level of homogeneous one when the water/air interface is replaced by an oil-water interface. Accordingly, we hypothesized that surface sealing of water with an appropriate oil phase could suppress primary heterogeneous ice nucleation at the surface and enable extended storage of deeply supercooled water.

First, we cooled a large ensemble of polystyrene tubes containing 1ml of ultra-pure water to -13 °C (Fig. 1(A-B)). This resulted in > 90% of samples to be frozen after 24 hours and nearly all samples to be frozen after 5 days. In contrast, the ultra-pure water samples could be kept in the liquid phase for a week, at the same temperature, if their surfaces were sealed by various types of immiscible oils, such as light mineral oil (MO), olive oil (OO), heavy paraffin oil (PO), and nutmeg oil (NO). Interestingly, the curdling of OO during DSC does not trigger water freezing, though the cumulative freezing frequency ($f_f$, $f_f = 1 - f_s$ ) increases significantly compared to water sealed by other oils (Fig. 1(A)). In supplementary experiments, we have observed that the water degassed by vacuuming for 24 hours, has similar $f_f$ as normal water, with or without oil sealing (Fig. S2). These experiments indicate that air dissolved in the water does not play a major role in ice nucleation in our experiments. Given this result and the consistent efficacy of surface sealing by different oils on freezing reduction, we infer that the air-water interface is the primary nucleation site.

We also examined the influence of water volume on the efficacy of oil sealing for freezing inhibition. We studied the two most promising oils, MO and PO, at -13 and -16 °C for ultra-pure water ranging from $10^0$ ~ $10^5$ µl (Fig. 1(C-D), Fig. S3). We found that MO sealing can effectively suppress water freezing for water volumes up to $10^4$ µl at -13 and -16 °C. PO sealing was even more effective with a low $f_f$ throughout the entire volume range at -13 °C, and only 45.8% of samples frozen at -16 °C for the $10^5$ μl samples. In addition, 8 out of 35 (22.8%) samples of $10^5$ μl water were kept in the supercooled state at -16 °C for 100 days without any freezing event after Day-3 (Fig. S3(B)). While further investigations might be necessary, these observations are incompatible with conventional stochastic freezing processes (SI), which implies exponential decrease of $f_s$ with time (*21, 22*). Alternatively, the freezing of DSC water sealed by oil could be depicted as "case-specific" that some of sealed water samples are more susceptible to crystallization than others due to their unique and slightly different microstructures on the interface, even though all of them possess higher capability to resist freezing compared to the water without sealing. A reconciliation of these two cases might lie in the fact that those samples that don't' freeze within our observation period have much fewer impurities and thus a much smaller exponential for the decay of $f_s$ than those that freeze within 3 days.

In order to further support our hypothesis that the water/air interface plays a dominant role in ice nucleation and subsequent freezing, we measured water freezing frequencies under differential degrees of surface sealing by MO, ranging from a) unsealed (0 oil), b) ring sealed along the contact line between water and tube wall (0.01 ml), c) partially sealed with partial exposure to air (0.1 ml), d) critically sealed with water surface just completely covered (0.15 ml), e) normally sealed (0.5 ml), and f) over sealed with excessive oil mounted on water surface (3.5 ml) (Fig. 2(A-B) and Fig. S4). The results indicate that the capacity of freezing inhibition increases with the degree of sealing, with a statistically maximum plateau achieved by critical sealing (Fig. 2(A)). Ring sealing (0.01 ml) that nullifies the triple solid/water/air contact line has a mild effect on freezing inhibition at high temperatures (-10 °C) but is not effective below -13 °C. Taken together with partial sealing results (0.1 ml), this result implies that the contact nucleation at the air-water-solid triple interface is not as dominant as that at water/air interface especially at low temperatures. Considering the crystallization efficiency depends on the integration of nucleation probability *J* and nucleation length (or area), the triple contact line of short length would provide smaller crystallization efficiency compared than the air/water interface even though it has higher *J* (*33, 34*). Overall, we confirmed that the water/air interface is the primary ice nucleation site for DSC water, and surface oil sealing that removes the water/air interface can effectively inhibit ice nucleation and water freezing.

We also observed that additional oil beyond the critical sealing has a statistically negligible effect on freezing suppression. This indicates that additional pressure and dampening effects, associated with a long-column of viscous oil phase, have a negligible effect on freezing inhibition. In order to further test this, we examined the effects of viscosity of the sealing agents where we used hydroxy (-OH) terminated polydimethylsiloxane (PDMS) of different chain lengths (Fig. 2(C)). In a similar fashion, we did not observe statistically significant differences in the capacity of freezing inhibition of PDMS with a viscosity range of 1 - $5x10^5$ cP, with the exception of 3500 cP PDMS which resulted in almost no freezing suppression effect. We hypothesize that this odd behavior is likely due to the formation of an ordered structure between water and this particular PDMS on the interface through hydrogen bonding, which closely matches the lattice of hexagonal ice (*35*).

Most oils are complex mixtures of alkanes, saturated cyclic alkanes, alkylated aromatic groups, and fatty acids among other hydrocarbon compounds. In an effort to more systematically study the observed freezing inhibition effect of supercooled water sealed with an immiscible hydrocarbon phase, we studied two prototypical families of hydrocarbons: linear alkanes and their corresponding primary alcohols of different lengths (Fig. 3). Specifically, we have studied alkanes

($C_mH_{2m+2}$, denoted $C_m$, m = 5 - 11) and primary alcohols ($C_mH_{2m+1}OH$, denoted $C_mOH$, m = 4 - 8) as the sealing agents for DSC water at -20 °C. Since linear alkanes have very low polarity, they have weak interaction with polar water molecules. On the other hand, the primary alcohols, which are amphipathic, can form strong hydrogen bonds with water through their hydroxyl group (hydrophilic end) and even stable ordered interfacial structures.

We found that $f_f$ of DSC water, at -20 °C, sealed with alkanes decreases monotonically with increasing carbon number m and chain length $l$ (Fig. 3(A)). The capacity of alkanes in freezing inhibition matches that of MO (Fig. 2(A) at -20 °C) at m > 9. Given that mineral oils tend to have carbon chain lengths above 10, this result is expected. We posit that this trend, correlated with the alkane chain length, might also explain the capacity differences between PO and MO in Fig. 1(D); PO likely consists of higher carbon chain alkanes than MO based on their densities (PO ~ 0.855-0.88 vs MO ~ 0.838 g/ml). On a molecular level, the mechanism for this trend might lie in the structure of the alkane/water interface. It has been observed that an interfacial electron depletion layer with a thickness $\delta$ exists between water and hydrophobic alkane chains by both X-ray reflectivity (XR) measurements (*36, 37*) and atomistic molecular dynamics (MD) simulations (*38, 39*). The few water molecules in the depletion layer (electron density < 40% that of bulk water (*40*)) can buckle in the intermolecular space near the ends of alkane molecules (Fig. 3(B)), and create a template for the formation of an ice nucleus (*29*). The alkane chains adjacent to the water molecules preferentially have their longest axis parallel to the water interface with a tilt angle $\beta$ (*36*). This tilt angle increases with m and $l$, resulting in a more parallel orientation for longer alkanes (*36*). Accordingly, longer alkane chains are expected reduce the corrugation and roughness of the interface on the side of alkanes. This, consequently, is expected to decrease the number of buckled water molecules and nucleus templates, and thus lower the probability of heterogeneous ice nucleation on that layer (*29*). These expectations are in line with our observations of decreasing freezing frequencies for longer alkane chain lengths. From the perspective of thermodynamics, longer and flatter-oriented alkanes results in fewer and sparser buckled water molecules in the interface serving as nucleation template, which implies smaller contact region between ice embryo and sealing alkanes, smaller $\theta_{iwo}$ (even though they are already much smaller than $\theta_{iwa}$), and higher energy barrier for heterogeneous ice nucleation (Fig. S1(B)).

On the other hand, $f_f$ of DSC water, at -20 °C, sealed with alcohols increases with m and $l$. This opposite trend is likely due to the starkly different structures of the alcohol/water interface compared to that of alkane/water interface (Fig. 3(C)). Unlike the alkanes which prefer a parallel orientation, the primary alcohols orient perpendicularly to the interface with a small $\beta$. The primary alcohols align their hydroxyl (-OH) heads toward the interface to form hydrogen bonds with water molecules. Accordingly, no depletion layer of interfacial water exists as in the alkane/water interface. The 2D layer of interfacial water molecules are strongly hydrogen-bonded to the hydroxyl groups, with their H atoms pointing toward alcohol as revealed by heterodyne-detected vibrational sum frequency spectroscopy (SFG) (*41*). Therefore, structures and dimensions of the contacting layer of amphilic alcohols essentially determine the distribution and arrangement of interfacial water molecules, and the formation of heterogeneous ice nucleus (*35, 42*).

Experimental and computational investigations of ice nucleation in droplets under monolayers of long primary alcohol chains with $16 \leq m \leq 31$, revealed a very low tilt angle $\beta$ (~ 7.5 - 10 °) and a very good lattice match between hexagonal ice and the alcohol structure for $29 \leq m \leq 31$ (*43*). For these longest chains ice nucleation occurs at temperatures as high as -1 °C. As m and $l$ decrease, the tilt angle $\beta$ increases up to ~ 19° for m = 16 (*44*). In conjunction, a greater lattice mismatch between hexagonal ice lattice and ordered alcohol layer at the interface along with a lower ice nucleation efficiency and freezing temperature were observed (*35, 43-45*). Extrapolating this information and trend to the shorter alcohols (4 ≤ m ≤ 8) in this study, we expect a larger tilt angle $\beta$ (e.g. $\beta$ = 28° for m = 6 and $\beta$ = 30° for m = 5 (*44*)) and a greater lattice mismatch between hexagonal ice and ordered alcohol structure given the general structural similarity of primary alcohols. Compared to longer alcohol chains, the interfacial -OH groups anchored to smaller alcohols have stronger in- and out-of-plane fluctuations at the same temperature. We, therefore, expect that the large lattice mismatch along with the –OH group fluctuations can destabilize any ordered domain of crystalline water and impede the formation of ice nucleus of critical size (*45*). Given that both effects are larger with smaller chain lengths, we expect that higher nucleation inhibition can be achieved by smaller primary alcohols, in line with our experimental observations. From the perspective of thermodynamics, greater lattice mismatch and interface fluctuation associated with shorter alcohol molecules directly reduce the probability of the formation of icing template of critical size for successful nucleation, which indicates smaller stable contact area between ice nucleus and sealing alcohols and thus, higher free energy barrier for heterogeneous ice nucleation at the interfac. Once again, we observed that there is no significant difference of $f_f$ between 1-day and 7-day storage when sealed by either alkanes or alcohols. This further suggests the case-specific, rather than stochastic, nature of water freezing with oil sealing that we have previously discussed.

Having established the efficacy of the deep supercooling approach using either oils or pure alkane and alcohol phases, we then studied its stability under vibrational, thermal, and ultrasonic disturbances. Vibrational disturbances were introduced by placing DSC water onto a shaking plate with various shaking speeds and frequencies (SI). When the DSC water (-20 °) is sealed by MO, its $f_f$ is 0% and 5.6%, respectively, under 0.84 g and 2.1 g centrifugal acceleration (Fig. 4(A)), which are are much higher than ac/deceleration forces of a commercial airliner (0.2-0.4 g) during potential transporation. Thermal distrubances were induced by putting the DSC samples into 37 °C incubator or plungig them into 37 °C water bath with warming rate of $10^0$ C/min (heated by natural convection in air) or $10^2$ C/min (heated by forced convection in water), respectively. Very few (0% for gas warming, 2.5% for water warming) of the samples freeze under these thermal fluctuations. In contrast, these samples can not endure ultrasonication in 40 kHz ultrasonic water bath (Fig. 4(A) and Video S1), with $f_f$ of ~ 84%. This is probably due to the vigorous collapse of cavitation bubbles in water during ultrasonication (*18*), which would cause ultrahigh local pressure (> 1 GPa) (*48*), and therefore, significantly increase equilibrium temperature $T_e$ and the degree of supercooling $\Delta T$.

Upon the instability of DSC water sealed by MO under ultrasonication, we further tested its stability sealed by pure alkanes and pimary alcohols. DSC water sealed by alkanes freeze immediately upon being ultrasonicated (Fig. 4(B-C) and Video S2), which is consistent with previous observation of MO sealed water since MO has a high content of various alkanes. On the contrary, none of the samples freeze upon ultrasonification if they were sealed by any of the primary alcohols (Fig. 4(B)). Instead, the sealing alcohols would be emulsified with supercooled water, starting from the interface and then evolving toward supercooled water (Fig. 4(C) and Video S3). The exact mechanism of the freezing resistance of DSC water sealed by alcohols to ultrasonic disturbance is still unknown, and one hypothesis would be that ultrasound preferrentially transduces its energy into joint molecular motion at interface due to the hydrogen bonding between water and amphilic alcohols to form nanoemulsion (*49*), rather than cavitation bubbles for ice nucleation in viscous DSC water.

In the preceding, we demonstrated a seemingly counterintuitive and novel approach to achieve long-term deep supercooling of large-volume water by using a hydrocarbon-based immiscible phase to seal the water surface. Our initial observations with laboratory grade oils demonstrated that replacing the water/air interface, which is the primary ice nucleation site, with a water/oil interface dramatically inhibits stochastic freezing processes. The seemingly time independent nature of the freezing frequency of oil-sealed water suggests that its freezing might be case-specific rather than stochastic. Our studies with linear alkanes and primary alcohols suggest that freezing inhibition can be achieved by surface sealing with starkly different interfacial structures and microscopic mechanisms, which results in opposite trends of inhibition capacity correlated to the chain length. While all sealed DSC water show great stability under vibrational and thermal disturbances simulating normal storage and transportation conditions, only the primary alcohol sealed supercooled water can withstand ultrasonication. We note that while we have hypothesized about possible macroscopic (thermodynamic) and microscopic mechanisms that might explain our observations, further studies are warranted to test, confirm, and improve upon them. Especially, since most existing literature focuses on longer alkane and alcohol chains at the water interfaces and the resulting molecular structures, computational and experimental studies with short chains might prove useful. Similarly, careful measurements of interfacial properties and structures at low temperatures with mixed and pure hydrocarbons can shed further light on why some oils are more effective than others.

Given that monolayers of long alcohol chains have been historically used to initiate ice nucleation, our results with the short chains to prevent nucleation expand the use of alcohols to provide a robust control mechanism over the temperature at which nucleation can be achieved in an aqueous solution. Further studies with different families of hydrocarbons and their mixtures will be aimed at expanding this robust control of supercooling degree to allow novel applications. Beyond its fundamental implications, deep supercooling of large volumes of aqueous solutions can enable previously prohibitive applications, and provide new biopreservation methodologies for cell, tissue, and organ engineering and transplantation, as well as other areas, such as food preservation. Given our prior experience and interest in both organ and cell preservation using supercooling and the limitations we have previously encountered in terms of temperatures, volumes and durations for preservation, we believe that the deep supercooling via the surface sealing with immiscible phases will be vital in advancing these applications forward. The immediate goal is to translate this approach to preservation of large number of cells that are amenable to preservation at the supercooled temperatures we can achieve here, and then translate such results to the clinic. We will then explore tissue and organ preservation with deep supercooling approaches.

## ACKNOWLEDGEMENTS

We are grateful to Prof. Mehmet Toner for the very helpful discussions and his suggestions. We would also like to thank the NIH for funding this work through grants no. 5P41EB002503 (BioMEMS Resource Center), 1R21EB020192, 5R01EB023812.

# Figures

Figure 1. Deep supercooling (DSC) of pure water enabled by surface sealing with various types of oils. (A) Cumulative freezing frequency ($f_f$) for 1 ml water at -13 °C over 7-day DSC, without sealing (W/O seal), with surface sealing by light mineral oil (MO), olive oil (OO), heavy paraffin oil (PO), and nutmeg oil (NO). Number of independent experiments n = 6, number of total tested samples for each case N = 56. NS: $p > 0.05$; *: $0.005 < p < 0.05$; **: $1.0 \times 10^{-6} < p < 0.005$, ***: $p < 1.0 \times 10^{-6}$. Error bar represents standard deviation. (B) Corresponding samples of (A) post 1-day storage. (C) $f_f$ of DSC water of various volumes post 1-day storage at -13 and -16 °C. n = 7, N = 272, 145, 336, 123, and 125 for 3, 30, 200, 1000, and 10000 μl water, respectively. (D) $f_f$ of 100000 μl water with different sealing oils and temperatures post 1-day storage, n = 7, N = 35.

Figure 2. Dependence of freezing efficiency of DSC water on the volume and viscosity of sealing agent. (A) Effect of sealing oil (MO) volume on $f_f$ post 1-day DSC at different temperatures. n = 6, N = 70. (B) Side view of corresponding samples of (A). MO includes Oil Red O for staining and imaging. 0 oil, 0.01 ml, 0.1ml, 0.15 ml, 0.5 ml, and 3.5 ml indicate no seal, ring seal, partial seal, critical seal (just complete surface seal), standard seal, and over seal by MO, respectively. (C) Effect of viscosity of sealing agents on $f_f$ post 1-day DSC at -16 °C. The sealing agents are hydroxy (OH) terminated polydimethylsiloxane (PDMS) of different chain lengths and viscosities. n = 5, N = 56.

Figure 3. Freezing efficiency of 1 ml water sealed with linear alkanes and primary alcohols at -20 °C. n = 7, N = 87. When m > 11 for linear alkanes and m > 8 for primary alcohols, the sealing agents are frozen at -20 °C and cause DSC water frozen. When m < 5, the linear alkanes are gaseous under atmospheric condition and not suitable for sealing. When m < 4, the primary alcohols are miscible with water and not suitable for sealing either. (B-C) Schematic configurations of alkane/water (B), and alcohol/water interface (C), respectively. The alkane and alcohol molecules are displayed without aliphatic hydrogen atoms and colored in light green. The O and H atoms in hydroxyl group of alcohol and water are shown in red and white dots, respectively.

Figure 4. Stability tests for -20 °C DSC water. (A) $f_f$ of DSC water sealed by MO under various disturbances. Vibrational disturbance was imposed by shaking plate with different shaking frequencies and centrifugal forces (i.e. 0.84 g or 2.1 g). Thermal disturbance was imposed by placing or plunging the DSC tubes into 37 °C incubator (37 °C gas) or water bath (37 °C water). Ultrasonic disturbance was introduced by putting the DSC tubes into 40 kHz ultrasonic water bath. n = 6, N = 48. (B) $f_f$ of DSC water sealed by linear alkanes and primary alcohols under 40 kHz ultrasonic disturbance. n = 3, N = 24 (except for $C_5$, N = 8). (C) Representative image sequences of ultrasonication tests for DSC water sealed by linear alkanes or primary alcohols.

Figure 1

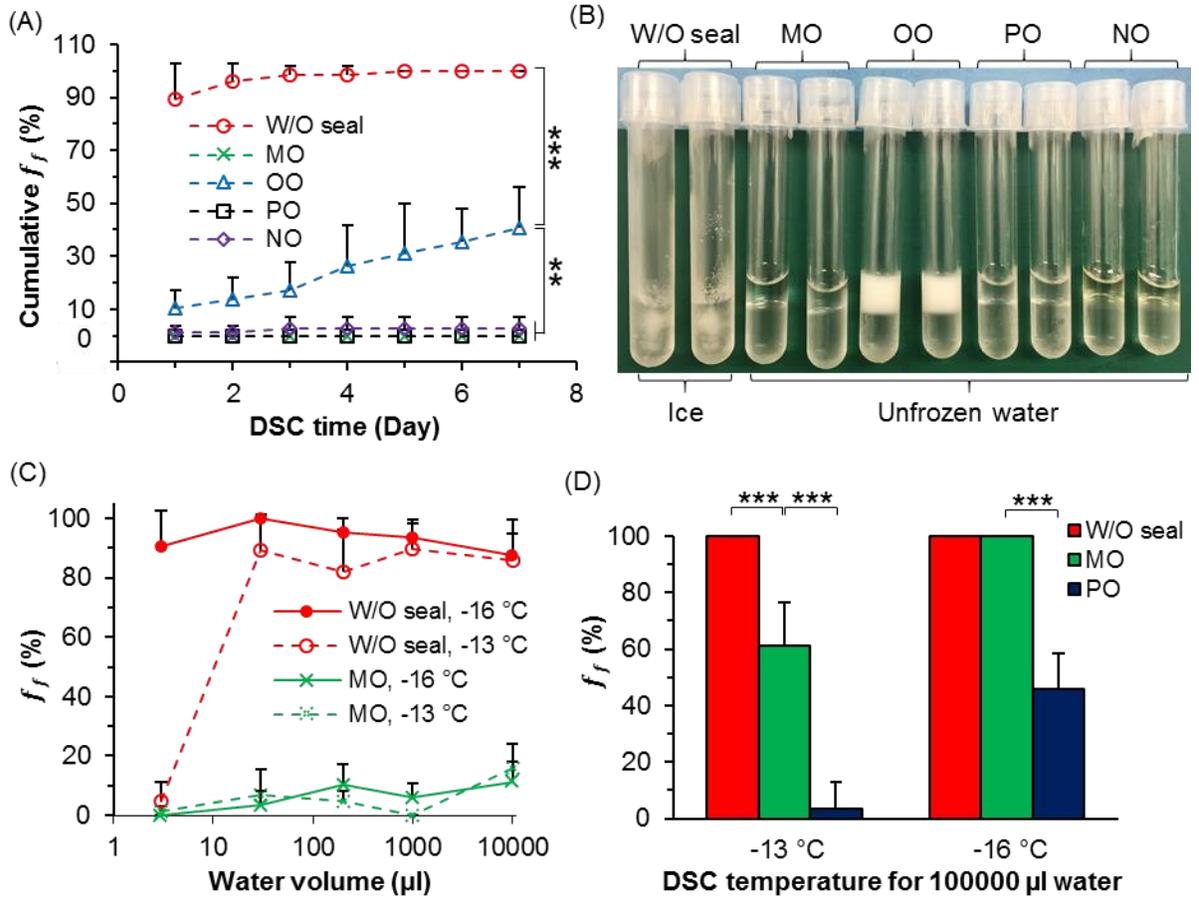

Figure 2

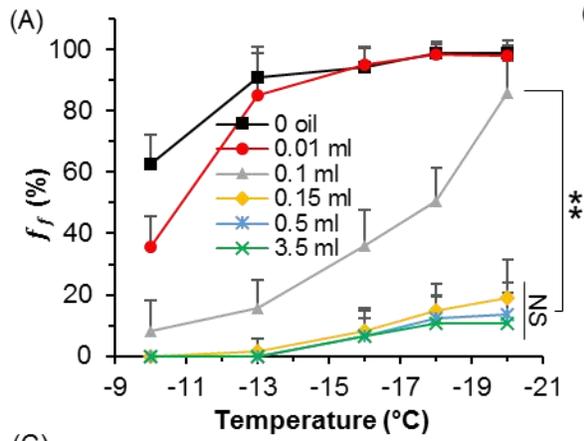
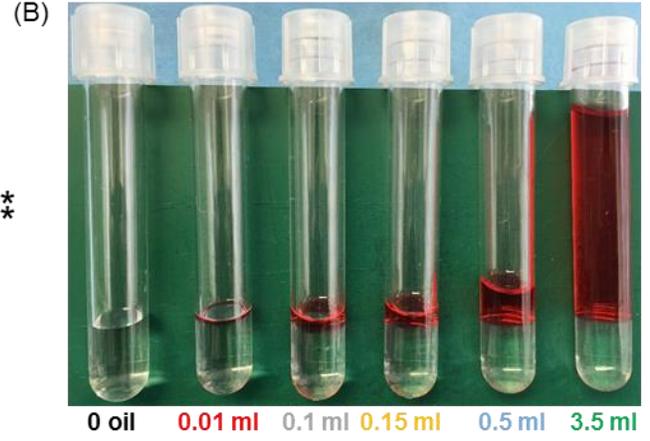
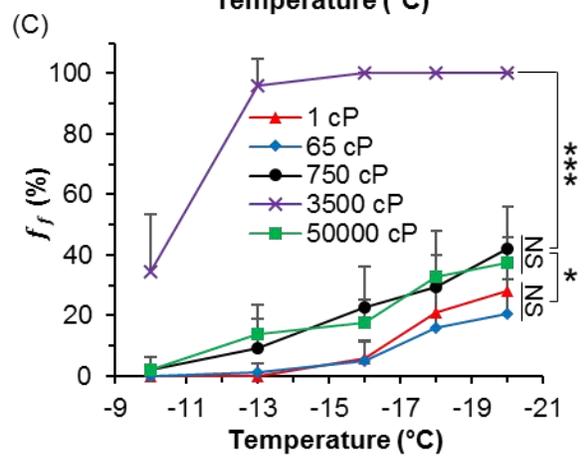

Figure 3

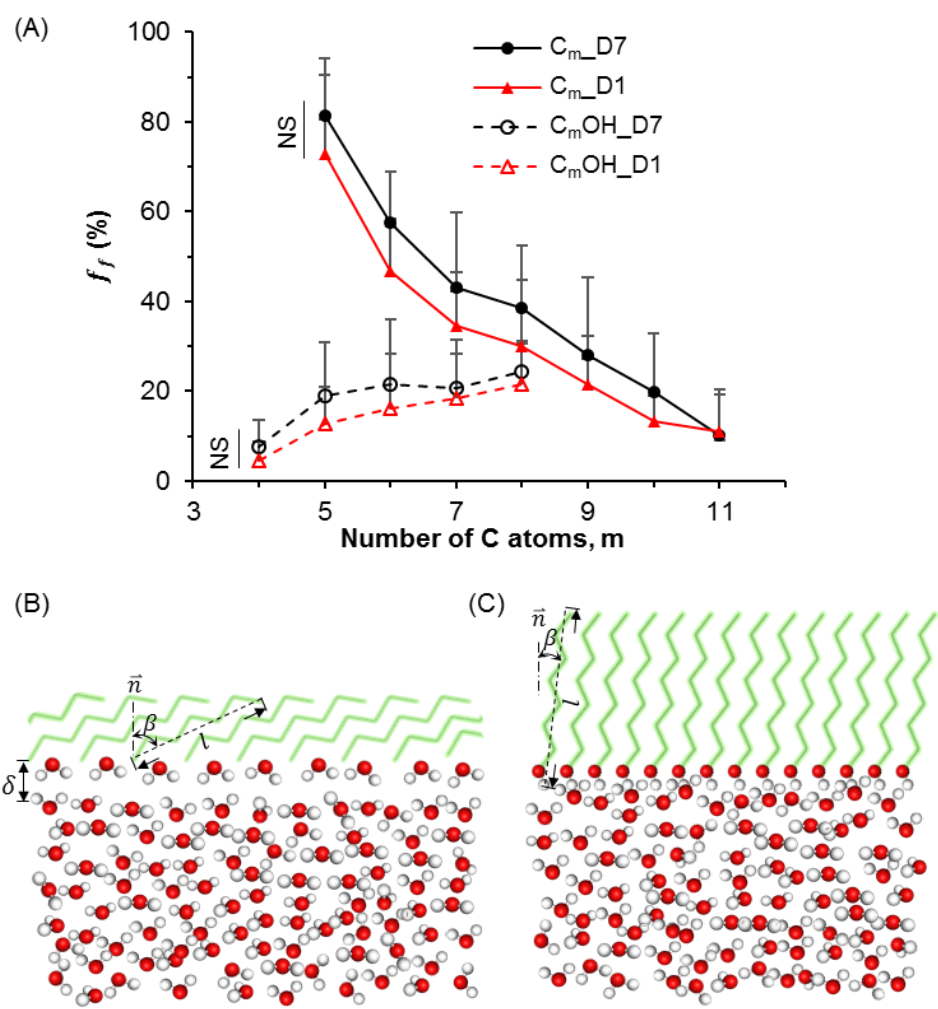

Figure 4

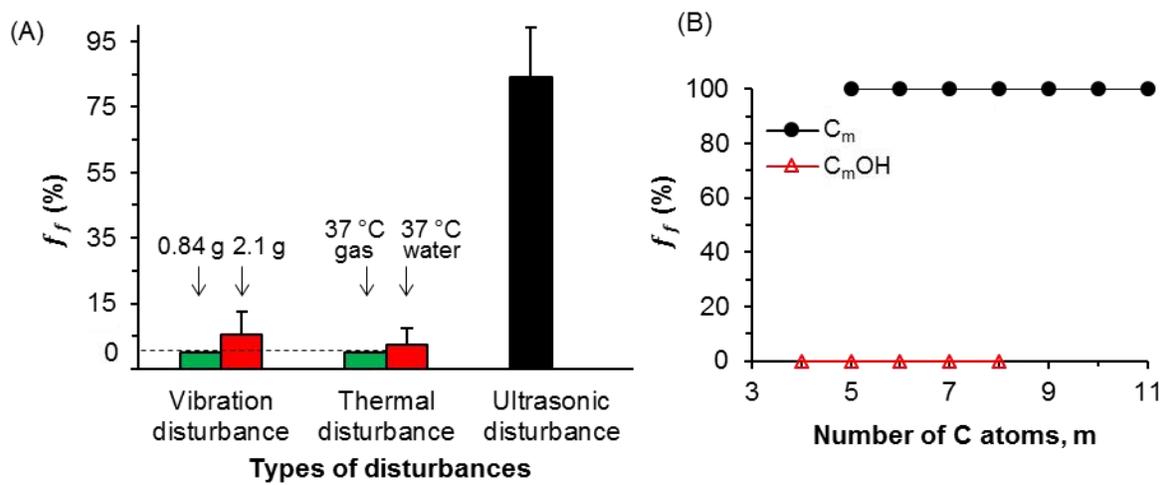

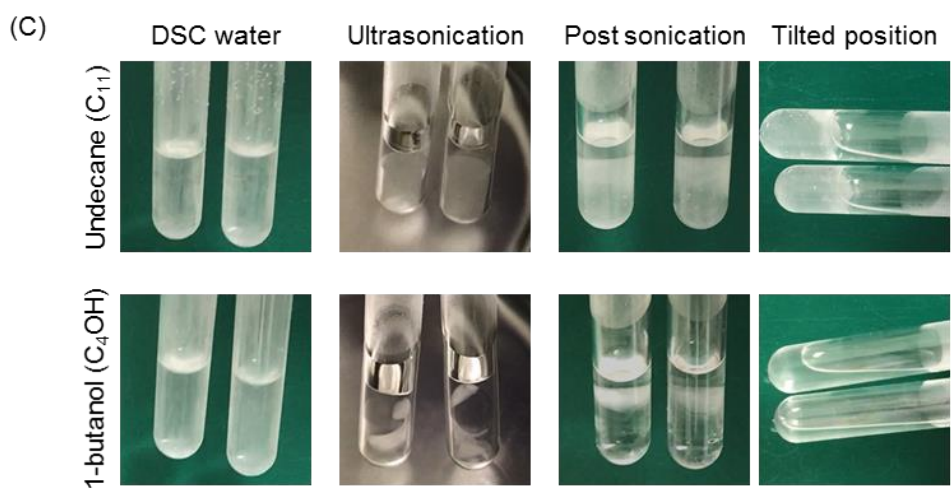

**Supplemental Information (SI)**

**Materials and Methods**

For all experiments in this study, DNase/RNase-free distilled water (Life Technologies/Thermo-Fisher Scientific, USA) was used to minimize potential pollutants or ice-nucleating agents, except DSC trials of 100 ml water where deionized (DI) water (resistivity $R = 18.2$ MΩ) produced by a deionizing water system (METTLER TOLEDO Thornton, USA) was used. All water containers (dishes, 96-well plates, round-bottomed tubes, and bottles, Corning, USA) were made of polystyrene, and clean and sterile before experiments. The purity of all oil phases used for water surface sealing is at least 99% as specified by the vendor (Sigma-Aldrich, USA).

The loading of water into containers was performed in a chemical hood to avoid contamination of the samples by pollutants or dust particles in the air. Water of small volume (< 1 ml) was loaded into containers (dishes or 96-well plates) using clean and sterile tips (Thermo Fisher Scientific, USA) and calibrated pipets (PIPETMAN, Gilson, USA), while that of large volume (≥ 1 ml) was loaded into containers (round-bottomed tubes or bottles) using serological pipets (Thermo Fisher Scientific, USA) by pipette filler (Drummond Scientific, USA). We note that as water droplets smaller than $10^0$ μl are subject to significant evaporation during long-term deep supercooling (DSC) experiments, and those bigger than $10^5$ μl (100 ml) beyond the volume capacity of the freezing chamber, they were not investigated in this study. After loading water samples into the containers, oil phase was gently added onto the water surface using serological pipets, trickling down along the wall of containers to avoid splashing or trapping air bubbles at the interface. The water-laden containers (with or without sealing oil) were transferred into portable temperature controlled freezers (Engel MHD-13, Engel, USA) that were placed in 4 °C cold room to minimize temperature fluctuations, or stored in -20 °C freezer (Thermo Fisher Scientific, USA). The temperatures within these freezers were verified by Toluene-filled low-temperature thermometer (Sigma, USA).

To examine the effects of dissolved air in water on ice nucleation and water freezing, the water was vacuumed at a pressure below $10^{-4}$ atmosphere for 24 hours to extract dissolved air molecules. The degassed water was, then, gently pipetted into tubes and sealed with mineral oil (MO) for supercooling tests at -16 °C. The air content of the degassed water is significantly lower than that of normal water without degassing, as no air bubbles emerge from the degassed water (second row of Fig. S2(B)) under vacuum. The same procedure was carried out for normal water for comparative purposes, and several big air bubbles can be observed after 3-hour degassing (first row of Fig. S2(B))

To test the stability of DSC water sealed by oil phase, three types of disturbances, vibrational, thermal, and ultrasonic disturbances, were studied. For vibrational disturbance test, DSC tubes were placed on shaking plate (Labline 4625 titer shaker, Marshall Scientific, USA) with shaking speed 500 and 800 rpm for 30 seconds, which give rise to the centrifugal acceleration of 0.84 and 2.1 g (g is gravitational acceleration), respectively. To prevent heat transfer, the tubes were wrapped with thick tissue paper in tube racks, all of which had been previously cooled to -20 °C in freezer. The temperature of the DSC water would not change noticeably during experiments given the brief shaking period and thick insulation layer. For thermal disturbance test, the DSC tubes were put into 37 °C incubator (warmed by air) or plunged into 37 °C water bath (warmed by water). Therefore, DSC water would experience different warming rates and temperature gradient. For ultrasonic disturbance test, DSC tubes would be plunged into 4 °C ultrasonic water bath. The sonicator (Branson B-200, TMC Inductries, USA) generates 40 kHz ultrasonic wave with power 30 W. The freezing of the DSC water can be evidenced by the change of sample transparency (from transparent to opaque).

All data were organized and reported as the mean ± standard deviation from at least three independent runs of experiments (n > 3); further information on sample numbers are disclosed in figure captions. The statistical significance of mean values between two groups was determined by Microsoft Excel based on Student's two-tailed *t*-test, assuming equal variance. Although a *p*-value less than 0.05 is generally regarded as statistically significant, different ranges of *p*-value (NS: 0.05 < *p*, *: 0.005 < *p* < 0.05, **: 0.005 < *p* < 10$^{-6}$, ***: *p* < 10$^{-6}$) were provided to show different degrees of significance.

**Stochastic process of ice nucleation and freezing**

The formation of a critical ice embryo, i.e. a successful nucleation, in metastable supercooled water is generally regarded as a stochastic process that does not depend on the number of previous nucleation trials or correlate to other nucleation events during the same period (*1, 2*). In addition, heterogenous nucleation is the major type of crystallization in this study since homogeneous nucleation in water occurs at much lower temperatures (around - 40 °C). As a result, the heterogeneous ice nucleation on water surfaces/interfaces would follow Poisson statistics

$$In\left(1 - f_f(t)\right) = -J(T) \cdot S \cdot t$$

where $f_f(t)$ is the freezing frequency after supercooling of a period $t$, $J(T)$ is the nucleation rate at temperature $T$, and $S$ is the area of heterogeneous nucleation sites. Therefore, for water samples of the same volume and shape under a constant temperature, the non-frozen (supercooled) fraction is expected to decline exponentially with time.

However, in our experiments we found that $f_f(t)$ of DSC water with oil sealing does not change significantly after Day 3 as shown in Fig. 1, Fig. 3, Fig. S2, and Fig. S3. These results indicate that the heterogeneous ice nucleation in DSC water sealed by oil phase does not follow the conventional theory of stochastic nucleation processes at the interface. Particularly, since 45.8% of 100 ml DSC water (-16 °C) sealed by PO are frozen post 1-day storage (Fig. 1(D)), $In\,(1 - 0.458) = -J(-16\,°C) \cdot S \cdot 1$, which gives rise to $J(-16\,°C) \cdot S$ = 0.612. Therefore, the expected fraction of unfrozen samples, $1 - f_f(t) = e^{-0.612t}$, would decrease exponentially with storage time for DSC water of the same volume and shape at -16 °C. As a result, the fraction of unfrozen samples would be 0.22% and 2.5×10$^{-25}$ % on Day-10 and Day-100, respectively, which implies almost all the sample would be frozen after 10-day storage. However, we observed that 22.9% (8 out of 35) of samples were still unfrozen post 100-day storage in our experiments. Moreover, no freezing event occurred between Day-3 and Day-100; that is all the samples that were unfrozen on Day-3 remained unfrozen till the end of our experiments on Day-100. These observations strongly demonstrated a non-stochastic process of ice nucleation in our DSC water-oil phase systems.

Using similar heuristics, our experimental results also suggest that the freezing we observed is not due to homogeneous ice nucleation either, where the formation of critical ice embryo is caused by spontaneous aggregation of water molecules via random translational, rotational, and vibrational movements that would conform stochastic process (*3*). The exact kinetics and statistics of this heterogeneous ice nucleation in DSC water sealed by oil phase is still unknown and detailed future investigations are certainly warranted.

**Freezing point depression due to oil-water mixing**

When a water sample is sealed by an oil phase (i.e mixed oils, and pure alkanes and alcohols) for DSC, the "immiscible" oil might slightly dissolve in supercooled water to decrease the equilibrium melting temperature $T_e$ below 0 °C, the equilibrium melting point of pure water under atmospheric conditions. We, therefore, quantified the potential depression of freezing point due to this effect and assessed whether it's comparable to the high degree of supercooling we observed in our experiments. According to the Bladgen's Law, the extent of freezing point depression $\Delta T_F$ can be calculated by

$$\Delta T_F = ibK_F$$

where $i$ is the Van't Hoff factor ($i$ = 1 for nonelectrolytes or oil phase in this study), $b$ is the molality of oil phase in water, and $K_F$ is the cryoscopic constant ($K_F$ = 1.85 K·kg/mol for water). Therefore, $\Delta T_F$ can be determined by the solubility of sealing oils in water at DSC temperatures. Solubility of oils in metastable water at DSC temperatures are not readily available; as such we have assessed $\Delta T_F$ using the available solubility data of oils in water under room temperature. This approach, likely, leads to an overestimation since solubility of oils in water typically increases with temperature (*4*).

For oil mixtures (MO, OO, PO, NO, and PDMS) and linear alkanes ($C_5$ ~ $C_{11}$) utilized in this study, the maximum solubility is 0.04 g/L (or 0.55 mM) ($C_5$ in water), and the corresponding estimate for $\Delta T_F$ is less than 1.03 × 10$^{-3}$ °C, which is negligible compared to the degree of supercooling $\Delta T$ (10 to 20 °C) achieved using these oil phases as sealing agents.

For alcohols used in this study, the maximum solubility is 73 g/L (or 0.98 M) ($C_4$OH in water at room temperature) and the corresponding estimate for $\Delta T_F$ is less than 1.82 °C. This likely overestimated freezing point depression accounts for about 9.1% of $\Delta T$ (20 °C) enabled by alcohol sealing. Moreover, the DSC water and sealing alcohols are likely not mixed altogether. The stable contact interface with strong hydrogen bonding on the head and a long hydrophobic tail of alcohols, low molecular mobility, and viscous water at - 20 °C would significantly impede the diffusion of alcohol molecules into water. We, therefore, conclude that the depression of freezing temperature due to oil-water mixing does not play a significant role in achieving the observed high degree of supercooling in our experiments.

**Supplemental figures**

Figure S1. Schematics for ice/water/air (A) and ice/water/oil (B) contacts. $θ_{iwa}$ and $θ_{iwo}$ are water contact angles on ice/water/air and ice/water/oil interfaces, respectively.

Figure S2. Comparison of cumulative freezing frequencies between normal and degassed DSC water at - 16 °C. "D_W/O seal", "D_MO", and "D_PO" represent degassed water without sealing, surface sealing with MO, and PO, respectively. n = 6, N = 60, NS: $p$ > 0.05. (B) Degassing images for normal and degassed water with or without MO sealing. The red dash circles indicate air bubbles precipitated under vacuum.

Figure S3. Representative images of DSC water of various volumes. (A) 30 µl droplets with or without MO sealing post 1-day DSC at - 16 °C. (B) 100 ml deionized (DI) water sealed by 16 ml PO post 100-day DSC at - 16 °C. All possible freezing events occur before Day-3, and 8 out of total 35 (n = 7, N = 35) bottles of water maintain unfrozen after 100-day storage.

Figure S4. Vertical view of water surface in round-bottomed tubes sealed by MO of various volumes. The oil was stained by Oil Red O for enhanced contrast. Newton's rings were observed due to the curvature of sealing oil near the tube wall.

**Supplemental videos**

Video S1. Ultrasonication for 1 ml DSC water sealed by MO. The water is supercooled at - 20 °C for 1 day and the ultrasonic frequency is 40 kHz.

Video S2. Ultrasonication for 1 ml DSC water sealed by undecane ($C_{11}$). The water is supercooled at - 20 °C for 1 day and the ultrasonic frequency is 40 kHz.

Video S3. Ultrasonication for 1 ml DSC water sealed by 1-butanol ($C_4OH$). The water is supercooled at - 20 °C for 1 day and the ultrasonic frequency is 40 kHz.

Figure S1

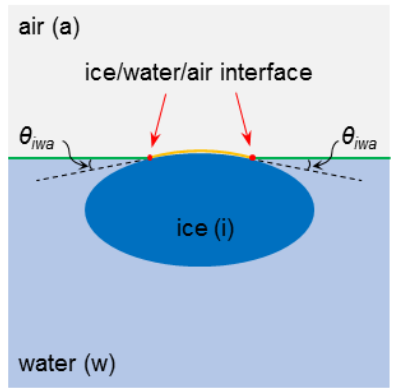 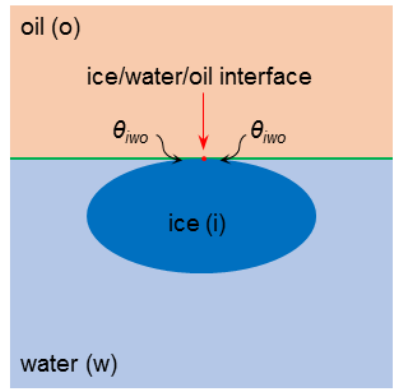

Figure S2

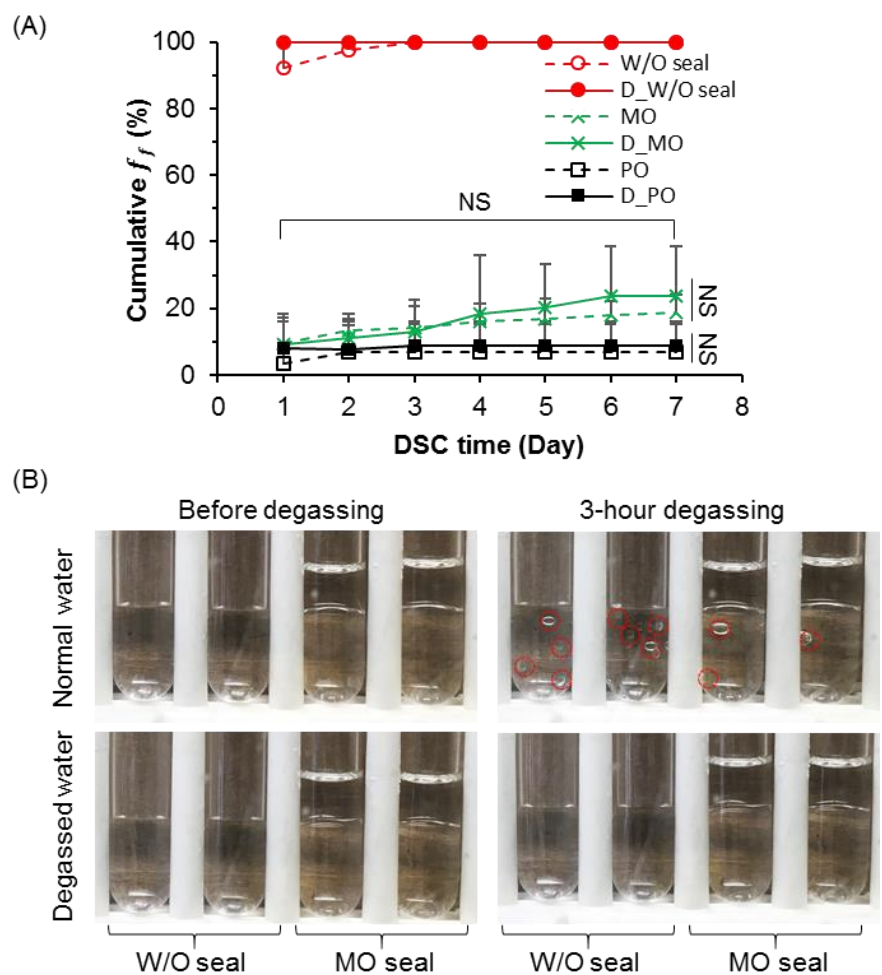

Figure S3

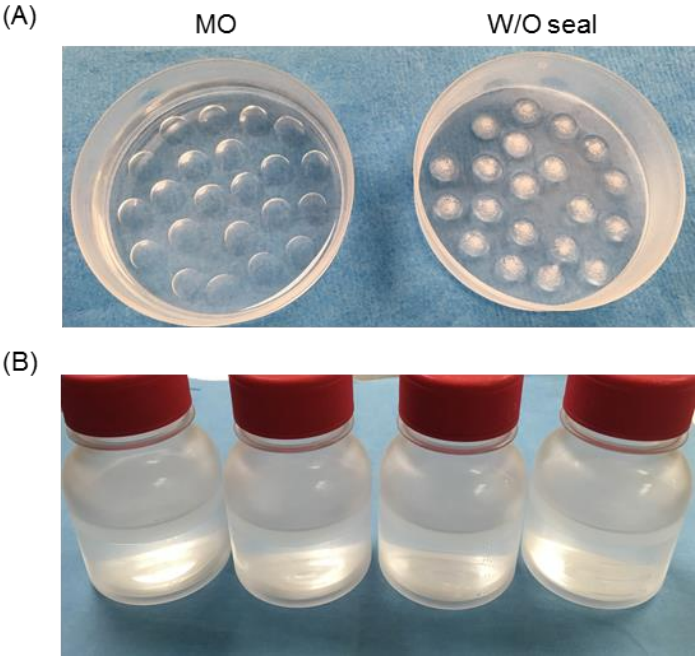

Figure S4

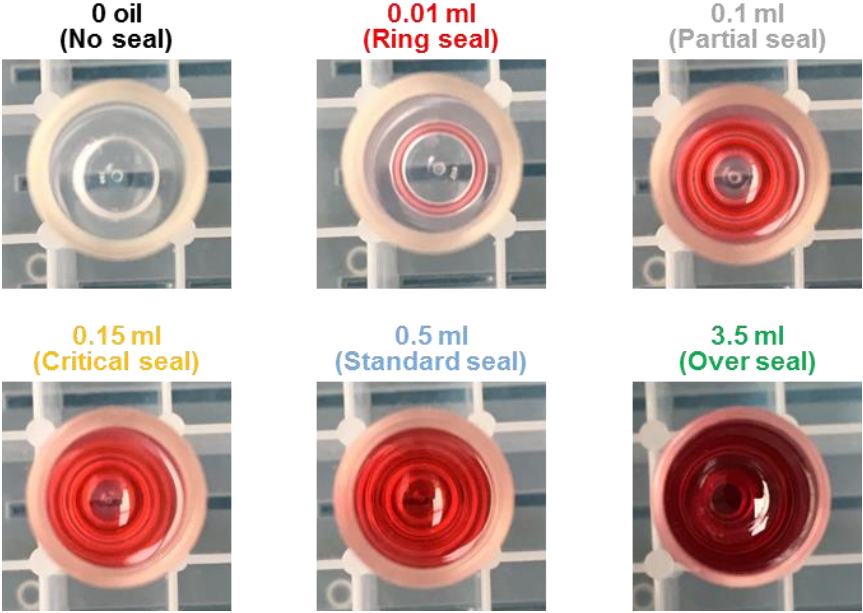